\documentclass[doublecol]{epl2}%
\usepackage{amsmath}%
\usepackage{amsfonts}%
\usepackage{amssymb}%
\usepackage{graphicx}
\providecommand{\U}[1]{\protect\rule{.1in}{.1in}}
\shorttitle{Spin critical opalescence in zero temperature Bose-Einstein Condensates}
\shortauthor{D. H. Santamore and E. Timmermans}
\institute{
  \inst{1} Department of Physics, Temple University - Philadelphia, PA 19122 USA\\
  \inst{2} T-4, Theory division, Los Alamos National Laboratory - Los Alamos, NM 87545
}
\abstract{Cold atom developments suggest the prospect of measuring scaling
properties and long-range fluctuations of continuous phase transitions at
zero-temperature. We discuss the conditions for characterizing the phase
separation of Bose-Einstein condensates of boson atoms in two distinct hyperfine
spin states. The mean-field description breaks down as the system approaches
the transition from the miscible side. An effective spin description clarifies the
ferromagnetic nature of the transition.  We show that a difference in the scattering
lengths for the bosons in the same spin state leads to an effective internal
magnetic field. The conditions at which the internal magnetic field vanishes
(i.e., equal values of the like-boson scattering lengths) is a special point.
We show that the long range density fluctuations are suppressed near that
point while the effective spin exhibits the long-range fluctuations that characterize
critical points. The zero-temperature system
exhibits critical opalescence with respect to long wavelength waves of impurity
atoms that interact with the bosons in a spin-dependent manner.}
\begin{document}

\title{Spin critical opalescence in zero temperature Bose-Einstein Condensates}
\author{D. H. Santamore\inst{1}
\and Eddy Timmermans\inst{2}}
\maketitle

\section{Introduction}

Now that cold atom technology has realized uniform trapping potentials bounded
by sharp edges~\cite{one,two}, the scaling of near
zero-temperature phase transitions can be explored in the laboratory. We
consider spin domain formation in a Bose-Einstein condensate (BEC) of atoms in
two distinct hyperfine states, which we call the ``spin up" ($\left|
\uparrow\rangle\right. $) and ``spin down" ($\left|  \downarrow\rangle\right.
$) states. The Feshbach tuning of one of the scattering lengths can trigger
this transition in the quantum (zero-temperature) regime. In-situ images of
the atoms in one of the spin states can reveal spin density fluctuations at a
characteristic length scale (the correlation length) that diverges as the
scattering length is tuned near a critical value. We show that the mean-field
description breaks down near the transition so that the critical exponents may
differ from their mean-field values. To provide a reference to future
experiments and to reveal trends with respect to polarization, scattering
lengths, and density, we calculate fluctuation properties in mean-field. Using
a spin description, we find that, for Ising spin-spin interactions with equal
$\uparrow- \uparrow$ and $\downarrow- \downarrow$ scattering lengths, the
spins exhibit long-range fluctuations whereas the long-range density
fluctuations are suppressed near the transition. The system then remains
transparent to distinguishable low energy atoms that interact with the bosons
in a spin independent manner whereas it turns opaque to atoms that experience
a spin-dependent interaction. This system can be realized with $^{87}Rb$ atoms
that support a resonance in the interaction of atoms in different
spin states~\cite{four,five}.

\section{Switching ground states}

The phase separation of boson superfluids was predicted in~\cite{ten}, its
cold atom realization, dynamics and surface tension were predicted and
described in~\cite{eleven} and the transition was demonstrated
in~\cite{twelve}. Here, we show that the ground state configuration of a
dilute homogeneous BEC of $N_{\downarrow}$ spin-down bosons and $N_{\uparrow}$
spin-up bosons of mass $m$ confined to a macroscopic volume $\Omega$ alters
from a homogeneous `miscible' mixture to an `immiscible' separated state.
Bosons at positions $\mathbf{x}$ and $\mathbf{x}^{\prime}$ interact via
effective short-range potentials $\lambda_{\downarrow(\uparrow)} \delta\left(
\mathbf{x}-\mathbf{x}^{\prime}\right) $ if they occupy the down (up) spin
state and via $\lambda_{U} \delta\left( \mathbf{x}-\mathbf{x}^{\prime}\right)
$ if their spins differ. The interaction strengths $\lambda$ are proportional
to the respective scattering lengths, $\lambda_{\uparrow,\downarrow,U}=(4
\pi\hbar^{2}/m) a_{\uparrow,\downarrow,U}$, one of which can be Feshbach-tuned
across the phase boundary.

In the immiscible ground state configuration, $N_{\downarrow}$ spin-down
bosons reside in a volume $\Omega_{\downarrow}$ and $N_{\uparrow}$ spin-up
bosons occupy $\Omega_{\uparrow} = \Omega-\Omega_{\downarrow}$. Assuming the
volumes are sufficiently large to neglect surface effects, the
immiscible ground state energy is
\begin{equation}
E_{sep} = \frac{ \lambda_{\downarrow} }{2 \Omega_{\downarrow}} N_{\downarrow
}^{2} + \frac{\lambda_{\uparrow}}{2 \left[  \Omega- \Omega_{\downarrow}
\right] } N_{\uparrow}^{2}.\label{esep}%
\end{equation}
We determine $\Omega_{\downarrow}$ by minimizing $E_{sep}$, which gives
$\lambda_{\downarrow} N_{\downarrow}^{2}/\left[  2 \Omega_{\downarrow}^{2}
\right]  = \lambda_{\uparrow} N_{\uparrow}^{2} / \left[  2 \left(
\Omega-\Omega_{\downarrow}\right) ^{2} \right] $, implying equal pressures in
$\Omega_{\downarrow}$ and $\Omega_{\uparrow}$. Inserting the corresponding
volume fraction, $\Omega/\Omega_{\downarrow} = 1 + (N_{\uparrow}%
/N_{\downarrow}) \sqrt{\lambda_{\uparrow}/\lambda_{\downarrow}}$ into
$E_{sep}$,
\begin{equation}
E_{sep}=\Omega_{\downarrow} \frac{\lambda_{\downarrow}N_{\downarrow}^{2}}{2
\Omega_{\downarrow}^{2}} + \left(  \Omega- \Omega_{\downarrow} \right)  \frac{
\lambda_{\uparrow} N_{\uparrow}^{2}}{2 \left( \Omega- \Omega_{\downarrow}^{2}
\right) ^{2}} = \Omega\frac{ \lambda_{\downarrow} N_{\downarrow}^{2}}{2
\Omega_{\downarrow}^{2}},\label{esepint}%
\end{equation}
we find that the separated ground state energy 
\begin{equation}
E_{sep}=\frac{\lambda_{\downarrow}}{2\Omega}N_{\downarrow}^{2}+\frac
{\lambda_{\uparrow}}{2\Omega}N_{\uparrow}^{2}+\frac{\sqrt{\lambda_{\uparrow
}\lambda_{\downarrow}}}{2\Omega}N_{\uparrow}N_{\downarrow}.\nonumber
\label{esep2}
\end{equation}
drops below the mean-field energy $E_{mix}$ of the homogeneous mixture,
\begin{equation}
E_{mix} = \frac{ \lambda_{\downarrow}}{2 \Omega} N_{\downarrow}^{2} +
\frac{\lambda_{\uparrow}}{2 \Omega} N_{\uparrow}^{2} + \frac{ \lambda_{U}
}{2\Omega} N_{\uparrow} N_{\downarrow},\label{emix}%
\end{equation}
if $\sqrt{\lambda_{\uparrow} \lambda_{\downarrow}} < \lambda_{U}$.
Characterizing the competition between like and unlike interactions by
\begin{equation}
g \equiv\frac{ \lambda_{U}^{2}}{\lambda_{\uparrow}\lambda_{\downarrow}} \; ,
\end{equation}
the BEC ground state switches at $g=1$.

\section{Divergence of single component compressibility}

We determine the spin-down density response $\delta\rho_{\downarrow}$ to a
potential perturbation $\delta V_{\downarrow}$. The low frequency, long
wavelength response follows from the Thomas-Fermi description that minimizes
the homogeneous free energy $F_{0}=E_{mix}-\mu_{\downarrow}N_{\downarrow} -
\mu_{\uparrow}N_{\uparrow}$ at fixed chemical potentials $\mu_{\downarrow}$
and $\mu_{\uparrow}$ and replaces $\mu_{\downarrow} \rightarrow\mu
_{\downarrow} - \delta V_{\downarrow}$. We solve the resulting equations
\begin{align}
\lambda_{\downarrow} \rho_{\downarrow} + \lambda_{U} \rho_{\uparrow}  & =
\mu_{\downarrow} - \delta V_{\downarrow}\nonumber\\
\lambda_{U} \rho_{\downarrow} + \lambda_{\uparrow} \rho_{\uparrow}  & =
\mu_{\uparrow} \;\label{tf}%
\end{align}
by linearizing $\rho_{\downarrow(\uparrow)}=\rho_{\downarrow(\uparrow),0} +
\delta\rho_{\downarrow(\uparrow) }$ around the homogeneous ($\delta
V_{\downarrow}=0$) equilibrium densities $\rho_{\downarrow(\uparrow),0}$. The
$\delta\rho_{\downarrow}$ result gives a response function (proportional to
the compressibility)
\begin{equation}
\chi_{\downarrow} = - \frac{\delta\rho_{\downarrow} } {\delta V_{\downarrow} }
= \frac{1}{\lambda_{\downarrow}} \frac{1}{[1-g]} \; ,\label{chi}%
\end{equation}
that diverges when $g \rightarrow1$. The divergence of the compressibility
implies large scale fluctuations that cause critical opalescence \cite{E} near
ordinary critical points. The contribution of $[-\left( \lambda_{\downarrow} g
\right) ]$ to the $\chi_{\downarrow}$-denominator of eq.~(\ref{chi}) describes
the long wavelength $\downarrow$--$\downarrow$ attraction mediated by the
$\uparrow$ BEC that competes with the short-range $\downarrow$-$\downarrow$
repulsion described by $\lambda_{\downarrow}$. The divergence implies that the
smallest of $V_{\downarrow}$-potential variations induces a large
$\rho_{\downarrow}$ density response. Equating the flat potential requirement
for simulating infinite systems to $\delta\rho_{\downarrow}/\rho
_{\downarrow,0}\ll1$ suggests that $V_{\downarrow}$ control should ensure
that
\begin{equation}
\Delta V_{\downarrow} \ll\mu_{\downarrow} [ 1 - g ],
\end{equation}
where $\Delta V_{\downarrow}$ represents the $\downarrow$ potential variation
over $\Omega$.

\section{Mean-field breakdown}

The mean-field description predicts the boundary of its validity
regime~\cite{fifteen}. The miscible ($g<1$) mean-field densities are
homogeneous, except for small amplitude fluctuations. This assumption breaks
down when $\delta\rho_{\uparrow}$-quantum fluctuations induce an effective
$\delta V_{\downarrow}=\lambda_{U} \delta\rho_{\uparrow}$-variation
sufficiently large to give $\left|  \delta\rho_{\downarrow} \right|  \sim
\rho_{\downarrow,0}$. Estimating the quantum fluctuation as $\left|
\delta\rho_{\uparrow} \right|  \sim\sqrt{ \lim_{r\rightarrow0} \langle
\delta\rho_{\uparrow} \left( \mathbf{{r} }\right)  \delta\rho_{\uparrow}
\left(  \mathbf{{0} }\right)  \rangle}$ so that $\left|  \delta\rho_{\uparrow}
\right|  / \rho_{0,\uparrow} \sim\left(  \rho_{0,\uparrow} a_{\uparrow}^{3}
\right) ^{1/4}$, we expect large $\delta\rho_{\downarrow}$ fluctuations when
\begin{equation}
\left[  1 - g \right]  < \left(  \frac{\lambda_{U}}{\lambda_{\downarrow}}
\right)  \left(  \frac{ \rho_{0,\uparrow} } {\rho_{0,\downarrow} } \right)
\left(  \rho_{0,\uparrow} a_{\uparrow}^{3} \right) ^{1/4} \; .
\end{equation}
where $a_{\uparrow}$ ($a_{\downarrow}$) is the scattering length of spin-up
(down) state. A similar relationship follows for $\delta\rho_{\uparrow}$.
Assuming similar densities and scattering lengths, $\rho_{\uparrow} \sim
\rho_{\downarrow} \sim\rho/2$, where $\rho$ is the total density, $\rho
=\rho_{\uparrow}+\rho_{\downarrow}$, $a_{\uparrow} \sim a_{\downarrow} \sim
a_{U} $ and introducing $a=\left(  a_{\uparrow} a_{\downarrow} a_{U} \right)
^{1/3}$ we suggest that
\begin{equation}
\left[  1 - g \right]  > \left( \rho a^{3} \right) ^{1/4} \; ,
\end{equation}
is a necessary condition for mean-field validity.

\section{Feshbach steering of the system and required magnetic-field control}

In a magnetically controlled Feshbach resonance it is the strength of a
homogeneous magnetic field $B$ near the resonant value $B_{0}$ that varies the
scattering length. How accurate does the magnetic field have to be controlled
to avoid the effects of a fluctuating interaction? The magnetic field
dependence of the scattering length
\begin{equation}
a=a_{0}\left[  1-\frac{\Delta}{B-B_{0}}\right] ,
\end{equation}
where $a_{0}$ denotes the background scattering length and $\Delta$ represents
the width, implies a scattering length variation, $\delta a$,
induced by a magnetic field variation $\delta B$ equal to
\begin{equation}
\delta a=a_{0}\left\vert \frac{\Delta}{B-B_{0}}\right\vert ^{2}\frac{\delta
B}{\Delta}.
\end{equation}
For the $^{87}Rb$ case, the near-equality of the triplet and singlet
scattering lengths leads to $a_{\uparrow} \simeq a_{\downarrow} \simeq
a_{\downarrow} \simeq5 nm$ within a few percent.
Hence B should be tuned far from resonance, $\left| B - B_{0} \right| \sim \Delta/\delta$ to
encounter the phase boundary.  This is important as the resonance of refs. \cite{four}, \cite{five}
has been found to quite 'lossy'.  At $\left| B - B_{0} \right| \sim 50 \Delta - 100 \Delta$, however, particle
loss should not play a role. This implies that the transition condition is
achieved far from the resonance with $\frac{\Delta}{B-B_{0}}=\delta$ a few
percent. The relative scattering length variation is, then, of order
$\delta^{2}$ even if $\left\vert \delta B\right\vert \sim\Delta$,
\begin{equation}
\left\vert \frac{\delta a}{a}\right\vert =\delta^{2}\left\vert \frac{\delta
B}{\Delta}\right\vert \; .
\end{equation}
Hence, it should be feasible to ensure that $\left|  \delta g/g \right|  = 2
\left|  \delta a_{U} / a_{U} \right|  << [1-g]$.

\section{Spin analogy}

\begin{figure}[ptb]
\onefigure{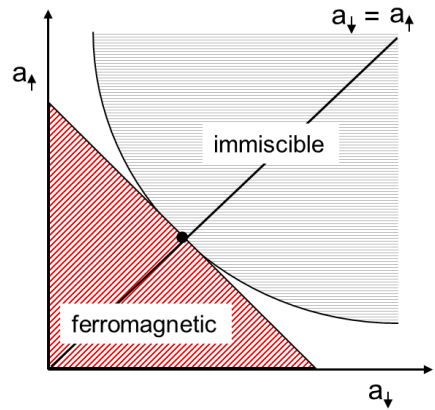}\caption{In spin-language, the difference of
$\uparrow\uparrow$ and $\downarrow\downarrow$ scattering lengths translates
into a short-range internal magnetic field carried by the particles
interacting with the other spins. This internal magnetic field can locally
align the effective ground state spins, even if the spin-spin coupling is
antiferromagnetic $\left(  \lambda_{L}\left(  1-r\right)  <0\right)  $. In
this figure, the $a_{\uparrow}$, $a_{\downarrow}$ phase diagram for fixed
$a_{u}$ value is shown. We illustrate the $\lambda_{\uparrow}-\lambda
_{\downarrow}$ effect on the zero-temperature transition by showing that the
immiscible regime (shaded by horizontal lines) and the ferromagnetic coupling
regime (shaded by diagonal lines) only intersect along the $a_{\uparrow
}=a_{\downarrow}$-line. Only when $a_{U}$ is varied while $a_{\uparrow
}=a_{\downarrow}$, does phase separation at zero temperature take place when
the spin coupling switches from ferro- to antiferromagnetic.}%
\label{fig.1}%
\end{figure}

Scaling exponents have been determined for finite temperature transitions of
spin lattices. We introduce the effective spin operator $\mathbf{{\hat{\sigma
}}}$ so that $\hat{\sigma}_{z}\left\vert \uparrow\right\rangle =+\left\vert
\uparrow\right\rangle $,
$\hat{\sigma}_{z}\left\vert \downarrow\right\rangle
=-\left\vert \downarrow\right\rangle $, and the $\hat{\sigma}_{x}$ and
$\hat{\sigma}_{y}$-operators are represented by the Pauli-matrices in the
$\left\vert \uparrow\right\rangle $, $\left\vert \downarrow\right\rangle
$--basis. The interaction of bosons with coordinates $\mathbf{r}$ and
$\mathbf{r}^{\prime}$, where $\mathbf{r}$ and $\mathbf{r}^{\prime}$ represent
both location $\mathbf{x}$ and $\mathbf{x}^{\prime}$ and spin $\mathbf{\hat
{\sigma}}$ and $\mathbf{\hat{\sigma}}^{\prime}$, is~\cite{sixteen}
\begin{align}
V_{eff}\left(  \mathbf{r},\mathbf{r}^{\prime}\right)   &  =\frac{\delta\left(
\mathbf{x}-\mathbf{x}^{\prime}\right)  }{4}\left[  \left(  \lambda_{\uparrow
}+\lambda_{\downarrow}+2\lambda_{U}\right)  \right. \nonumber\\
&  +\left(  \lambda_{\uparrow}-\lambda_{\downarrow}\right)  \left(
\hat{\sigma}_{z}+\hat{\sigma}_{z}^{\prime}\right) \nonumber\\
&  + \left.  \left(  \lambda_{\uparrow}+\lambda_{\downarrow}-2\lambda
_{u}\right)  \hat{\sigma}_{z}\hat{\sigma}_{z}^{\prime}\right] .
\label{spinint}%
\end{align}
The term linear in $\hat{\sigma}_{z}$ describes an effective short-range
magnetic field carried by the particles. This effective field interacts with
the other spins. The corresponding interaction term contributes a mean-field
energy that is indistinguishable from that of an effective magnetic field
$\mathbf{h}_{eff}$, $\mathbf{h}_{eff} \cdot\hat{\mathbf{\sigma}}$, where
$\mathbf{h}_{eff} = \left[ \left(  \lambda_{\uparrow} - \lambda_{\downarrow}
\right) /4\right]  \rho\hat{z}$. Characterizing the interactions by
\begin{equation}
\lambda_{L} = \frac{\lambda_{\uparrow} + \lambda_{\downarrow}}{2} \; , \; \; d
= \frac{ \lambda_{\uparrow} - \lambda_{\downarrow}}{\lambda_{\uparrow}%
+\lambda_{\downarrow}} \; \; , \; \mathrm{and} \; \; \; \; r = \frac
{\lambda_{U}}{\lambda_{L}},
\end{equation}
the spin-spin interaction potential reads
\begin{align}
V_{eff}(\mathbf{r},\mathbf{r}^{\prime})  &  =\delta\left(  \mathbf{x}%
-\mathbf{x}^{\prime}\right)  \frac{\lambda_{L}}{2}\left[  \left(  1+r\right)
\right. \nonumber\\
&  +\left.  d\left(  \hat{\sigma}_{z}+\hat{\sigma}_{z}^{\prime}\right)
+\left(  1-r\right)  \left(  \hat{\sigma}_{z}\hat{\sigma}_{z}^{\prime}\right)
\right] ,
\end{align}
$g$ takes the form $g= r^{2}/[1-d^{2}]$ and $\mathbf{h}_{eff} = (d/2)
\lambda_{L} \rho\hat{z}$. The zero internal magnetic field condition, $d=0$,
provides a special point: only for $\mathbf{h}_{eff}=0$ do the ground state
spins locally align when the spin-spin coupling term turns ferromagnetic as
illustrated in fig.~\ref{fig.1}. The spin analogy suggests that $\langle
\hat{\sigma}_{z} \rangle$ plays the role of the order
parameter\footnote{Measurements of $\langle\hat{\sigma}_{z} \rangle$ in the
$\Omega_{\uparrow}$ and $\Omega_{\downarrow}$-volumes would record the
different branches of the magnetization curve. At finite temperature,
$\langle\hat{\sigma}_{z} \rangle$ could vary discontinuously across the
transition as a function of $T$, except for $P=0$ at $d=0$ -- the critical
point. In analogy with finite temperature phase separation transitions, we
suggest that there may be a line (plotted as a function of P) of spinodal
decomposition and a coexistence line that touch at the critical point. The
second-order nature of the zero-temperature transition (independent of $P$)
may be a consequence of these lines approaching each other as $T \rightarrow
0$. At $P=0$, $d=0$, and fixed temperature value, the $\langle\hat{\sigma}_{z}
\rangle$ -variation as a function of $r$ tends to its asymptotic value $\pm1$
over an $r$-interval of magnitude $k_{B} T / \left[  \lambda_{L} \rho\right]
$, so that for $T=0$, $\langle\hat{\sigma}_{z} \rangle$ 'jumps' from $0$ to
$\pm1$, but this quantity still provides a legitimate order parameter.},
possibly offset by its value in the homogeneous mixture. 

\section{Zero temperature itinerant ferromagnet-like transition while avoiding
``lower branch" physics}

The effective spin description reveals that the transition combines
ingredients of phase separation and ferromagnetic transitions \cite{Onuki}.
The fermion analogue, discussed in the pioneering work on quantum phase
transitions~\cite{six}, was reported in a cold atom
trap~\cite{seven}. That transition, however, requires strong repulsion as the
components separate when the inter-particle interaction energy outweighs the
kinetic energy. In this strong coupling regime, two-particle bound states
(dimers) form involving `lower branch'
physics~\cite{seven.b,eight,nine.b,nine,nine.a,nine.c}. In contrast, BEC-phase
separation is triggered by the competition of different short-range
interactions, all of which can be weak, allowing the system to remain in its
metastable BEC state.

\section{Mean-field fluctuations}

We describe the correlations induced by the quantum fluctuations of the normal
modes that diagonalize the free energy operator
\begin{equation}
\hat{F}=\hat{H}-\mu_{\uparrow}N_{\uparrow} -\mu_{\downarrow} N_{\downarrow},
\end{equation}
where $\hat{H}$ denotes the Hamiltonian. The zero-momentum replacement of the
creation and annihilation operators $\hat{c}_{j,\mathbf{k}=0}$, $\hat
{c}_{j,\mathbf{k}=0}^{\dagger}$ $\rightarrow\sqrt{N_{j}}$ where $j=\uparrow
,\downarrow$ leads to the above mean-field free energy
\begin{equation}
F_{0}=E_{mix}-\mu_{\uparrow}N_{\uparrow} -\mu_{\downarrow} N_{\downarrow}.
\end{equation}
The next order contribution in the Bogoliubov expanded free energy $\hat
{F}=\delta\hat{F} + F_{0}$ keeps the quadratic terms
\begin{align}
\delta\hat{F}  &  =\sum_{\mathbf{k},j=\uparrow,\downarrow}\left[  e_{k}\left(
\hat{c}_{j,\mathbf{k}}^{\dagger}\hat{c}_{j,\mathbf{k}}\right)  \right.
\nonumber\\
&  +\left.  \lambda_{j}\rho_{0,j}\left(  \frac{\hat{c}_{j,\mathbf{k}}%
^{\dagger}+\hat{c}_{j,-\mathbf{k}}}{\sqrt{2}}\right)  \left(  \frac{\hat
{c}_{j,-\mathbf{k}}^{\dagger}+\hat{c}_{j,\mathbf{k}}}{\sqrt{2}}\right)
\right] \nonumber\\
&  +\sum_{\mathbf{k}}2\lambda_{U}\sqrt{\rho_{0,\uparrow}\rho_{\downarrow}%
}\left(  \frac{\hat{c}_{\uparrow,\mathbf{k}}^{\dagger}+\hat{c}_{\uparrow
,-\mathbf{k}}}{\sqrt{2}}\right)  \left(  \frac{\hat{c}_{\downarrow
,-\mathbf{k}}^{\dagger}+\hat{c}_{\downarrow,\mathbf{k}}}{\sqrt{2}}\right)  ,
\label{df1}%
\end{align}
where $e_{k}=\hbar^{2}k^{2}/2m$, and the chemical potentials,
$\mu_{\uparrow}$, $\mu_{\downarrow}$, were replaced by eq.~(\ref{tf}) with
$\delta V_{\downarrow} = 0$.
Two-component BEC Bogoliubov transformations were constructed
long ago, but these treatments obscure the underlying oscillator-structure \cite{Larsen}, \cite{Bassichis}.
As in Ref. \cite{sixteen} we introduce the phase space position/momentum-like operators
\begin{equation}
\hat{\phi}_{j,k}=\frac{\hat{b}_{j,k}^{\dag}+\hat{b}_{j,-k}}{\sqrt{2}};\pi_{j,k}=\frac{\hat{b}_{k}^{\dag}-\hat{b}_{-k}}{i\sqrt{2}},j=\uparrow,\downarrow
\end{equation}
that map $\delta F$ onto a sum of oscillator Hamiltonians. Below, the $k-$subscript and the sum over
repeated indices will be tacitly understood.  Unlike Ref. \cite{sixteen}, we make use of the
simplectic approach to phase space transformations \cite{Goldstein}.
We denote the phase space vector $\zeta\equiv\left(
\phi_{\uparrow},\phi_{\downarrow},\pi_{\uparrow},\pi_{\downarrow}\right)  $
and write $\delta F=(1/2) \zeta_{i}K_{ij}\zeta_{j}$ where\begin{equation}
\mathbf{K\equiv}\left(
\begin{array}
[c]{ll}\mathbf{K}_{\phi} & 0\\
0 & \mathbf{K}_{\pi}\end{array}
\right)
\end{equation}
\begin{equation}
\mathbf{K}_{\phi}\equiv\left(
\begin{array}
[c]{ll}e+\lambda_{\uparrow}n_{\uparrow} & \lambda_{u}\sqrt{n_{\uparrow}n_{\downarrow
}}\\
\lambda_{u}\sqrt{n_{\uparrow}n_{\downarrow}} & e+\lambda_{\uparrow}n_{\uparrow}\end{array}
\right)  ,\mathbf{K}_{\pi}\equiv\left(
\begin{array}
[c]{ll}e & 0\\
0 & e
\end{array}
\right) \; .
\end{equation}
The Bogoliubov transformation is a linear,
canonical point transformation $\zeta\rightarrow\eta$, $\eta\equiv\left(  \phi
_{+},\phi_{-},\pi_{+},\pi_{-}\right)  $ and $\zeta_{i}=M_{ij}\eta_{j}$ where
\begin{equation}
\mathbf{M}\equiv\left(
\begin{array}
[c]{ll}
\mathbf{M}_{\phi} & 0\\
0 & \mathbf{M}_{\pi}
\end{array}
\right)  ,
\label{M-matrix}
\end{equation}
resulting in a transformed $\mathbf{K}$-matrix of the form
\begin{equation}
\mathbf{K}^{\prime}=\mathbf{\tilde{M}KM}=\left(
\begin{array}
[c]{ll}
\mathbf{E} & 0\\
0 & \mathbf{E}
\end{array}
\right)  ,\label{K-matrix}
\end{equation}
where $\mathbf{\tilde{M}}$ represents the transpose of $\mathbf{M}$ and
$\mathbf{E}$ is the diagonal matrix of collective mode eigenvalues
\begin{equation}
\mathbf{E}\equiv\left(
\begin{array}
[c]{ll}
E_{+} & 0\\
0 & E_{-}
\end{array}
\right)  .
\end{equation}
The eigenvalue equations (\ref{M-matrix}) and (\ref{K-matrix}) take the form
\begin{eqnarray}
\mathbf{E}  & =\mathbf{\tilde{M}}_{\phi}\mathbf{K}_{\phi}\mathbf{M}_{\phi},
\label{Ematrix phi}\\
\mathbf{E}  & =\mathbf{\tilde{M}}_{\pi}\mathbf{K}_{\pi}\mathbf{M}_{\pi}.
\label{Ematrix pi}
\end{eqnarray}
The $\mathbf{K}^{\prime}$ is then equivalent (up to a constant) to the diagonalized
Bogoliubov Hamiltonian $\delta F^{\prime}=\sum_{\sigma=\pm}E_{\sigma}
b_{\sigma}^{\dag}b_{\sigma}$ where $b_{\sigma}^{\dag}=\left(  \phi_{\sigma
}+i\pi_{\sigma}\right)  /\sqrt{2}$, $b_{\sigma}=\left(  \phi_{\sigma}
-i\pi_{\sigma}\right)  /\sqrt{2}$. To ensure that $b_{\sigma},b_{\sigma}
^{\dag}$ satisfy the boson commutator relations, the $\mathbf{M}$
transformation has to be canonical \cite{Goldstein} implying
\begin{equation}
\mathbf{M}_{\mathbf{\phi}}\mathbf{\tilde{M}}_{\pi}=\mathbf{M}_{\mathbf{\pi}}\mathbf{\tilde{M}}_{\phi}=\mathbf{I}
\label{M-properties}\end{equation}
Writing $\mathbf{M}_{\mathbf{\phi}}$ as the product of a rotation matrix
$\mathbf{R}$\ ($\mathbf{\tilde{R}}=\mathbf{R}^{-1}$) and a diagonal scaling
matrix $\mathbf{S}$, we have $\mathbf{M}_{\mathbf{\phi}}=\mathbf{RS}$ with\begin{equation}
\mathbf{R}=\left(
\begin{array}
[c]{ll}\cos\theta & -\sin\theta\\
\sin\theta & \cos\theta
\end{array}
\right)  ,\mathbf{S}=\left(
\begin{array}
[c]{ll}\Gamma_{+} & 0\\
0 & \Gamma_{-}\end{array}
\right)  .
\end{equation}
We satisfy Eq.\ (\ref{M-properties}) by choosing $\mathbf{M}_{\mathbf{\pi}}=\mathbf{RS}^{-1}$. As $\mathbf{K}_{\pi}=e\mathbf{I}$, Eq.\ (\ref{Ematrix pi}) leads to $\mathbf{S}^{-2}=\mathbf{E}/e$ or $\Gamma_{\pm}=\sqrt{e/E_{\pm}}$.
Inserting the corresponding $\mathbf{S}$ into Eq.\ (\ref{Ematrix phi}) results
in\begin{equation}
\mathbf{\tilde{R}K}_{\mathbf{\phi}}\mathbf{R}=\mathbf{S}^{-1}\mathbf{ES}^{-1}=\left(
\begin{array}
[c]{ll}\mathbf{E}_{+}^{2}/e & 0\\
0 & \mathbf{E}_{-}^{2}/e
\end{array}
\right)
\end{equation}
so that $\mathbf{E}$\ follows from the diagonalization of $\mathbf{K}_{\mathbf{\phi}}$ and $\mathbf{E}_{+}^{2}/e$ and $\mathbf{E}_{-}^{2}/e$ are
the eigenvalues of $\mathbf{K}_{\mathbf{\phi}}$.
For a mixture of overall polarization, $P=\left(  N_{\uparrow}-N_{\downarrow
}\right)  $, we cast the resulting transformation in terms of the average
sound velocity $\bar{c}=\sqrt{\lambda_{L}\rho/m}$, the length $\xi
=\hbar/\left(  m\bar{c}\right)  $ and a transition parameter\begin{equation}
t=\left(  1-g\right)  \frac{\left(  1-d^{2}\right)  \left(  1-P^{2}\right)
}{(1+dP)^2}.
\end{equation}
Diagonalizing the $\mathbf{K}_{\mathbf{\phi}}$, we find\begin{equation}
E_{\pm}^{2}=\hbar^{2}k^{2}c_{\pm}^{2}\left(  1+\xi_{\pm}^{2}k^{2}\right)  ,
\end{equation}
where\begin{eqnarray}
c_{\pm}  & =\bar{c}\sqrt{1+dP}\sqrt{\frac{1\pm\sqrt{1-t}}{2}},\\
\xi_{\pm}  & =\frac{\bar{\xi}}{\sqrt{1+dP}}\sqrt{\frac{1}{1\pm\sqrt{1-t}}}.
\end{eqnarray}
The $\mathbf{S}$ matrix elements, $\Gamma_{\pm}$, take the form $\Gamma_{\pm
}=S\left(  \xi_{\pm}k\right)  $ with $S\left(  x\right)  =\sqrt{\frac{x^{2}}{1+x^{2}}}$ and the rotation angle $\theta$ in $\mathbf{R}$ is
determined by
\begin{equation}
\cos\left(  \theta\right)  =\sqrt{ \frac{1}{2} \left[  1+\left(  \frac{d+P}{1+dP}\right)
\frac{1}{\sqrt{1-t}}\right]  }.
\end{equation}
The corresponding transformation $\phi_{\uparrow\left(  \downarrow\right)
}\rightarrow\Phi_{+\left(  -\right)  }$, $\pi_{\uparrow\left(  \downarrow
\right)  }\rightarrow\Pi_{+\left(  -\right)  }$ with $\left\langle \Pi
_{\sigma,k}^{\dag}\Pi_{\sigma^{\prime},k^{\prime}}\right\rangle =\left\langle
\Phi_{\sigma,k}^{\dag}\Phi_{\sigma^{\prime},k^{\prime}}\right\rangle
=(1/2)\delta_{\sigma,\sigma^{\prime}}\delta_{k,k^{\prime}}$ determines all
correlation and response functions.
For instance, the mean-field $\uparrow\uparrow$-density correlation function
is
\begin{equation}
\langle\hat{\rho}_{\uparrow}\left(  \mathbf{x}\right)  \hat{\rho}_{\uparrow
}\left(  0\right)  \rangle=\rho_{0,\uparrow}^{2}+\rho_{0,\uparrow}\int
\frac{d^{3}k}{\left(  2\pi\right)  ^{3}}e^{i\mathbf{k}\cdot\mathbf{x}}\left[
2\langle\hat{\phi}_{\mathbf{k},\uparrow}\hat{\phi}_{-\mathbf{k},\uparrow
}\rangle\right]
\end{equation}
where%
\begin{align}
\left[  2\langle\hat{\phi}_{\mathbf{k},\uparrow}\hat{\phi}_{-\mathbf{k}%
,\uparrow}\rangle\right]  =s\left(  \xi_{+}k\right)  \cos^{2}\left(
\theta\right)  +s\left(  \xi_{-}k\right)  \sin^{2}\left(  \theta\right) .
\end{align}
Note that the correlation functions harbor two length scales: $\xi_{+}$ and
$\xi_{-}$. As $\xi_{-}$ diverges, $\xi_{-} \approx t^{-1/2}\overline{\xi}
\sqrt{2/\left(  1 + dP\right) ^{2}}$, this length scale, the correlation
length, exceeds the imaging resolution before the transition is reached.

For the special case of Ising spin-spin interactions, $d=0$, we determine the
long-range part of the density-density and spin-spin correlation functions,
\begin{equation}
\langle\hat{\rho}\left(  \mathbf{x}\right)  \hat{\rho}\left(  \mathbf{0}%
\right)  \rangle\approx\rho_{0}^{2}+\rho_{0}\;\int\frac{d^{3}k}{\left(
2\pi\right)  ^{3}}e^{i\mathbf{k}\cdot\mathbf{x}}s\left(  \xi_{-}k\right)
A_{\rho\rho}^{-},
\end{equation}%
\begin{equation}
\langle\hat{\sigma}_{z}\left(  \mathbf{x}\right)  \hat{\sigma}_{z}\left(
\mathbf{0}\right)  \rangle\approx\langle\hat{\sigma}_{z}\rangle^{2}+\rho
_{0}\;\int\frac{d^{3}k}{\left(  2\pi\right)  ^{3}}e^{i\mathbf{k}%
\cdot\mathbf{x}}s\left(  \xi_{-}k\right)  A_{\sigma_{z},\sigma_{z}}^{-},
\end{equation}
where $\rho_{0}=\rho_{0,\uparrow}+\rho_{0,\downarrow}$ and where the $A^{-}%
$-amplitudes depend on the Bogoliubov angle, giving%

\begin{align}
A_{\sigma_{z}\sigma_{z}}^{-}  &  =\frac{1}{4}\left(  \sqrt{1-P}\sqrt
{1+\frac{P}{\sqrt{1-t}}}\right. \nonumber\\
&  +\left.  \sqrt{1+P}\sqrt{1-\frac{P}{\sqrt{1-t}}}\right)  ^{2}%
\end{align}%
\begin{align}
A_{\rho\rho}^{-}  &  =\frac{1}{4}\left(  \sqrt{1-P}\sqrt{1+\frac{P}{\sqrt
{1-t}}}\right. \nonumber\\
&  -\left.  \sqrt{1+P}\sqrt{1-\frac{P}{\sqrt{1-t}}}\right)  ^{2}%
\end{align}
Hence, the near-transition spin-spin correlation function exhibits the
long-range order parameter correlations typical of a second-order phase
transition. In contrast, the long-range part of the density-density
correlations are suppressed, $A_{\rho\rho}^{-}\sim t^{2}$ if $P \neq0$ near
the transition. Even though the mean-field approximation breaks down, we
suggest that the near-transition suppression of long-range density
fluctuations is an actual feature of $d=0$ fluctuations.

\begin{figure}[ptb]
\onefigure{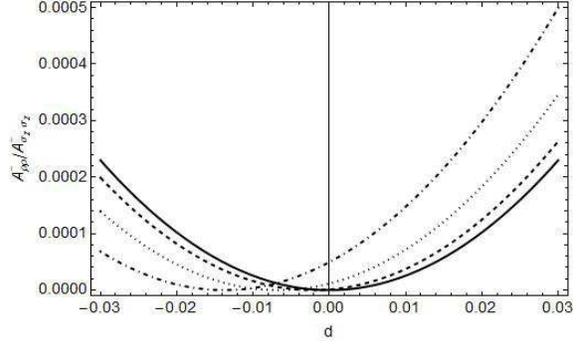}\caption{Plots of the ratio of the long-range amplitudes
of the zero-temperature density-density and spin-spin correlation functions as
a function of the relative difference $d$ of the $\uparrow\uparrow$ and
$\downarrow\downarrow$ scattering lengths, $d=\left[  a_{\uparrow} -
a_{\downarrow} \right]  / \left[  a_{\uparrow} + a_{\downarrow} \right]  $,
for fixed coherence length $\xi_{-} = 10 \overline{\xi}$ and different values
of the polarization, $P= \left[  N_{\uparrow}-N_{\downarrow} \right]  / N$,
with $P=0$ (full line), $P=0.2$ (dashed line), $P=0.5$ (dotted line) and
$P=0.7$ (dash-dotted line).}%
\label{fig.2}%
\end{figure}

The correlation functions can be extracted from the pixel counts of a single
in-situ $\uparrow$-density image, averaging $\rho_{\uparrow} \left(
\mathbf{R}+\mathbf{x}/2 \right)  \rho_{\uparrow} \left(  \mathbf{R} -
\mathbf{x}/{2}\right) $ over $\mathbf{R}$ (making the self-averaging
assumption that the $\mathbf{R}$--average is equivalent to averaging over many
samples). In accordance with the suppression of long-range density
fluctuations, we suggest that $\rho_{\uparrow}$ for a single image can be
extracted from the $\uparrow$-density. A $\rho_{\uparrow}$ image that resolves
features on the $\xi_{-}$ length scale though not on the $\xi_{+}$ scale can
be converted into a $\rho_{\downarrow}$ image by $\rho_{\downarrow}\approx
\rho_{0}-\rho_{\uparrow}$. In the case of $^{87}Rb$, $d$ nearly vanishes but
is not exactly zero ($d\sim0.01$). Therefore, we investigate the effect of a
small but finite $d$-value on the ratio of the long-range density-density and
spin-spin correlation amplitudes in fig.~\ref{fig.2}. In this figure, we plot the $A_{\rho\rho}^{-}/A_{\sigma_{z}%
\sigma_{z}}^{-}$ ratio as a function of $d$ for different $P$-polarizations
when $\xi_{-}=10 \overline{\xi}$. Note that for $d$ ranging from $-0.03$ to
$0.03$, the $A_{\rho\rho}^{-}/A_{\sigma_{z}\sigma_{z}}^{-}$-ratio remains
smaller than one part in one thousand with $P$ ranging up to $0.7$, suggesting
that the near transition suppression of the long-range density fluctuations
remains valid even for small but finite $d$-values.

\section{Spin opalescence}

In a classical system in equilibrium at temperature T, the long wavelength structure factor is
proportional to $k_{B} T \kappa$ where $\kappa$ denotes the isothermal compressibility.  Near the critical
point of a gas-liquid transition, the paradigm of critical opalescence, $\kappa$ diverges.  The
compressibility is also the long wavelength limit of the static density response function.
In the ground state of the spin1/2 BEC system with Ising spin interactions (d=0), it is the
spin response that diverges, not the density response.  To show that, we consider a weak magnetic field
perturbation of good momentum $\mathbf{k}$ described by a contribution $\delta H e^{i {\mathbf k}
\cdot {\mathbf x}} \Sigma_{z}({\mathbf x})/\Omega$ to the energy density where $\Sigma_{z}({\mathbf x}) =
\left[ \hat{\psi}_{\uparrow}^{\dagger}({\mathbf x}) \hat{\psi}_{\uparrow}({\mathbf x}) -
\hat{\psi}_{\downarrow}^{\dagger}({\mathbf x}) \hat{\psi}_{\downarrow}({\mathbf x})  \right] / \Omega$ to
the energy density.  The long-range part of the mean-field static response
$\delta \Sigma_{z} ({\mathbf x}) = \delta \Sigma_{z} e^{i {\mathbf k}
\cdot {\mathbf x}}/\Omega$ is given by $\delta \Sigma_{z} = \chi_{\sigma_{z}\sigma_{z}}^{-} \delta H$ where
\begin{equation}
\chi_{\sigma_{z}\sigma_{z}}^{-} = A_{\sigma_{z}\sigma_{z}}^{-} \frac{\rho}{mc_{-}^{2}} \frac{1}{1+ k^{2} \xi_{-}^{2}} \; ,
\label{spinresp}
\end{equation}
which diverges $\propto t^{-1}$ at the transition, whereas the analoguous long-range density
response function $\chi_{\rho\rho}^{-} \propto A_{\rho\rho}^{-}/c_{-}^{2} \propto t$ vanishes near the transition.
As a consequence, the spin fluctuations can mediate interactions in a very pronounced manner
and the Hamiltonian terms neglected in the Bogoliubov approximation become important.

\section{Feasibility and summary}

Critical slowing prevents an actual zero-temperature crossing of a
second-order phase transition: the temperature $T$ should be lower than an
energy scale, $mc_{-}^{2}$, that vanishes at the transition ($mc_{-}^{2} \sim
t$ in mean-field).
Experimentally, it would be very interesting to reach the mean-field
breakdown regime where the fluctuations acquire large amplitudes,
under the conditions of experimental access: $k_{B}T < mc_{-}^{2}$, the measuring time is
$\tau_{M} \gg \hbar/[mc_{-}^{2}]$, the potential variation $\Delta V$ remains sufficiently
small, and the system's linear size $L$ significantly exceeds the coherence length.
For a $^{87}Rb$ two-component BEC of density $\rho\sim 5\times10^{13} cm^{-3}$, $P\sim0.1$,
$a \sim 5 nm$, we estimate that mean-field breaks down at $g \sim 0.95$, where
$\xi\sim5 \overline{\xi}$.  The above values give $L\gg5\overline{\xi}$, $k_{B} T < mc_{-}^{2}/20$,
$\tau_{M}\gg 20 \hbar/[m\overline{c}^{2}]$ and $\Delta V < m\overline{c}^{2}/20$.
These conditions, while challenging, can be met in experiments.

We have discussed the prospect of characterizing the dilute gas BEC phase
separation of bosons in distinct hyperfine states as a zero-temperature
second-order phase transition. For equal like-boson scattering lengths, we
expect the system to exhibit long-range spin-spin correlations whereas the
long-range density fluctuations are suppressed.

\acknowledgments
ET's work was funded by the LDRD-program of Los Alamos National Laboratory. We
thank Malcolm Boshier for helpful comments.

\end{document}